\documentclass[twocolumn]{apex}
\usepackage{subfigure}

\usepackage{amsmath}
\usepackage[abs]{overpic}
\usepackage{color}

\usepackage{soul}

\title{Two-step pulse observation for Raman-Ramsey coherent population trapping atomic clocks}

\author{Yuichiro Yano$^1$, Shigeyoshi Goka$^1$, and Masatoshi Kajita$^2$}
\inst{$^{1}$Graduate School of Science and Engineering, Tokyo Metropolitan University, Minami-Osawa,
Hachioji, Tokyo 192-0397, Japan\\
$^{2}$National Institute of Information and Communications Technology, Koganei, Tokyo 184-8795, Japan\\ E-mail: yano-yuichiro@ed.tmu.ac.jp}
\abst{
We propose a two-step pulse observation method to enhance frequency stability for coherent population trapping (CPT) atomic clocks.
The proposed method is a Raman-Ramsey scheme with low light intensity at resonance observation, and provides a Ramsey-CPT resonance with both reduced frequency sensitivity to the light intensity and a high signal-to-noise ratio by reducing the repumping into a steady dark state.
The resonance characteristics were calculated based on density matrix analysis of a $\Lambda$-type three level system that was modeled on the $^{133}$Cs-D$_1$ line, and the characteristics were also measured using a vertical-cavity surface-emitting laser and a Cs vapor cell.
}

\begin{document}
\maketitle

Atomic clocks based on coherent population trapping (CPT) resonance have attracted attention as a means of fabricating very small frequency references, such as chip-scale atomic clocks\cite{knappe2004microfabricated}.
CPT atomic clocks are in great demand for many applications, such as telecommunications,
navigation systems, and synchronization of networks\cite{vig1992military}, and such clocks are required for their high frequency stability.

Frequency stability in atomic clocks is generally represented by the Allan deviation.
The frequency stability can be classified into two types based on the averaging time of the Allan deviation, which are short-term and long-term frequency stability; these two frequency stability types are degraded by different mechanisms.
To enhance the short-term frequency stability, a high signal-to-noise ratio (SNR) and a narrow CPT resonance linewidth are required because the short-term stability is determined by the product of the SNR and the Q value.
For long-term frequency stability, the light shift (or the ac Stark shift), which is a frequency shift caused by a light field, is a major limiting factor\cite{kozlova2014limitations}.
Because the long-term stability degradation induced by the light shift is caused by laser light power fluctuations or aging of the optical elements, reduced frequency sensitivity to the light intensity is desired.

The Raman-Ramsey (RR) scheme for enhancement of the frequency stability of CPT atomic clocks has been investigated by a number of researchers\cite{zanon2005high,esnault2013cold,xi2010coherent,yoshida2013line,yano2014theoretical}.
This scheme significantly reduces the variation in the light shift to be one or two orders of magnitude lower than that under continuous wave (cw) illumination\cite{castagna2009investigations,yano2014theoretical,yoshida2013line}.
Also, because this scheme suppresses power broadening, a narrow linewidth can be obtained\cite{zanon2005high}.
Recently, a short-term stability of 3.2$\times 10^{-13}\tau^{-1/2}$ was obtained, and stability as low as 3$\times10^{-14}$ was achieved at an averaging time of 200 s using a Cs vapor cell with the RR scheme\cite{danet2014dick}. 
Also, in the case where a cold Rb atom was used, short-term stability of 4$\times 10^{-11}\tau^{-1/2}$ was obtained, and a long-term frequency stability of 3$\times 10^{-13}$ was achieved for an averaging time of 5 h \cite{donley2014frequency}.

In our previous paper, we investigated the light shift in the RR scheme both theoretically and experimentally with the aim of enhancing the long-term frequency stability of CPT atomic clocks\cite{yano2014theoretical}.
The results showed that the light shift with the RR scheme is significantly reduced when compared with that under cw illumination.
We also found that the frequency variation of the light intensity is reduced by setting the observation time to be as short as possible.
This main reason for this is that the atoms evolve towards a steady state during observation.
In addition, we found that the light shift in the RR scheme is a nonlinear function of the light intensity,
and reduced frequency sensitivity is obtained under a high light intensity beyond a set threshold value.
Thus, the measurement conditions with both short observation times and high light intensity provide reduced frequency sensitivity to the light shift.
However, the short observation time causes the short-term stability to degrade because of increases in both the Dick effect and the laser intensity noise\cite{danet2014dick}.
In addition, while the high light intensity provides a high transition population, it also leads to a reduction of the resonance signal because of increased repumping into the steady dark state.\cite{guerandel2007raman,zanon2011ultrahigh}
Therefore, it is difficult to maintain the SNR of the resonance while simultaneously reducing the variation in the light shift.

In this paper, we propose a two-step pulse observation method for enhanced frequency stability in CPT atomic clocks.
The two-step pulse method is an RR scheme with low light intensity at observation.
This method provides low variation in the light shift without reducing the SNR.
We describe a light shift of the 6$^2$S$_{1/2}$ ($F=3\leftrightarrow 4$) state of $^{133}$Cs as observed through CPT using the two-step pulse observation method.
The light shift with the RR scheme is calculated based on the density matrix.
We also propose an estimation equation for the light shift with the RR scheme.
This equation is useful for estimation of the light shift with the two-step pulse observation method.

\begin{figure}[]
\centering
\includegraphics[width=3in]{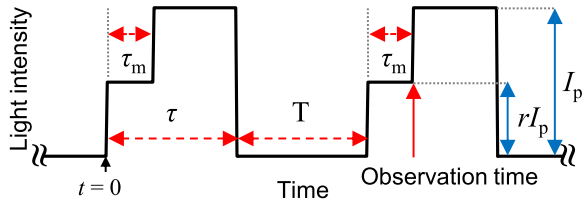}
\caption{(Color online) Ramsey pulse sequence: $\tau $ is the excitation time, $T$ is the free evolution time, and $\tau_m$ is the resonance signal observation time.}
\label{fig:scheme}
\end{figure}

The observation scheme for the two-step pulse observation method is shown in Fig. \ref{fig:scheme}.
The Ramsey-CPT resonance is observed by measuring the transmitted light intensity at the observation time $\tau _m$ after the pulse rise.
The two-step pulse observation maintains low light intensity for $rI_p$ until the observation point.
$r$ is defined as the observation intensity ratio, with a range of values from 0 to 1.
When $r=1$ is set, the two-step pulse observation method is treated with the same scheme as that used for the conventional RR scheme.
After the measurement is taken, the light intensity is changed into $I_p$ and the atoms are prepared for the next measurement.
A laser pulse with duration of ($\tau-\tau_m$) irradiates the atoms for pumping of the steady dark state.
After the free evolution time $T$, a laser pulse irradiates the atoms again with light intensity $rI_p$.
The light intensity $I(t)$ for this scheme is written as
\begin{equation}
I(t)=\left\{
\begin{array}{l l}
rI_p &(0<t\leq\tau_m)\\
I_p&(\tau_m<t\leq\tau)\\
0&(\tau<t\leq\tau+T)\\
\end{array}
\right..
\label{eq:scheme}
\end{equation}

\begin{figure}[]
\centering
\begin{overpic}[scale=0.5]{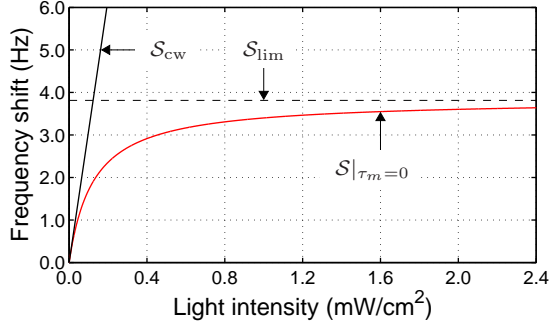}
\put(50,100){\colorbox{white}{\footnotesize$\mathcal{S}_{\rm cw}$}}
\put(85,100){\colorbox{white}{\footnotesize$\mathcal{S}_{\rm lim}$}}
\put(120,55){\colorbox{white}{\footnotesize$\mathcal{S}|_{\tau_m=0}$}}
\end{overpic}
\caption{(Color online) Light shifts as a function of the light intensity: $\mathcal{S}_{\rm cw}$ is the light shift under cw illumination, 
$\mathcal{S}|_{\tau_m=0}$ is the light shift under the RR scheme with conditions of $\tau_m =0$ and $\tau \to \infty$, and ${S}_{\rm lim}$ is the limit of $\mathcal{S}|_{\tau_m=0}$ as the light intensity approaches infinity.}
\label{fig:lightshift}
\end{figure}
\noindent

The calculated light shift in the RR scheme as a function of the light intensity is shown in Fig. \ref{fig:lightshift}.
The D$_1$-line of Cs was used as the excitation transition, and the 6$^2$P$_{1/2}$, $F=3$ state was selected as the excited state.
The calculation method is shown in more detail in Ref. \citen{yano2014theoretical}.
The black solid line is the calculated light shift under cw illumination.
The light shift under cw illumination is proportional to the light intensity ($\mathcal{S}_{\rm cw}=\alpha_{\rm cw} I_p$).
$\alpha_{\rm cw}$ is the slope of the light shift.
The red line is the calculated light shift with the RR scheme in the absence of the influence of observation $\mathcal{S}|_{\tau_m=0}$.
The free evolution time was set at 800 $\mu$s, and the $\tau \rightarrow \infty$ steady state solution is used as the initial condition to determine the evolution at later times\cite{zanon2011ultrahigh}.

The light shift in the RR scheme is a nonlinear function of the light intensity, and this light shift has two characteristics.
The first characteristic is that the variation under the RR scheme is equal to that under cw illumination with low light intensity.
The reason for this is that the effective light field applied under the RR scheme is the same as that applied under cw illumination with low light intensity, because the low light intensity leads to a long coherence time.
The second characteristic is that the light shift under the RR scheme approaches a saturation value $\mathcal{S}_{\rm lim}$ as the light intensity $I_p$ approaches infinity.
This saturation value $\mathcal{S}_{\rm lim}$ is inversely proportional to the free evolution time $T$ ($\mathcal{S}_{\rm lim}=C_{\rm lim}/T$).
When taking both of these characteristics into account, the light shift estimation equation in the absence of an observation time, $\mathcal{S}|_{\tau_m=0}$, is found to be as follows:
\begin{equation}
\mathcal{S}|_{\tau_m=0}=\frac{\alpha_{\rm cw}I_p}{\alpha_{\rm cw}C_{\rm lim}^{-1} TI_p+1}.
\label{eq:lightshift_tau0}
\end{equation}
Under low light intensity conditions ($I_p \ll C_{\rm lim}/(\alpha_{\rm cw}T)$), because the first term of the denominator is negligible, the value of the light shift $\mathcal{S}$ approaches that produced under cw illumination.
Also, under high light intensity conditions ($I_p \gg C_{\rm lim}/(\alpha_{\rm cw}T)$), because the second term of the denominator is negligible, the light shift $\mathcal{S}$ converges on $C_{\rm lim}/T$ $(=\mathcal{S}_{\rm lim})$.
Therefore, Eq. ({\ref{eq:lightshift_tau0}}) is satisfied using the light shift characteristics above, under the RR scheme.

Under the influence of observation ($\tau_m >0$),
the light shift under the RR scheme then depends on $\tau_m$.
For large $\tau_m$, the light shift under the RR scheme approaches that produced under cw illumination.
Because the light intensity at observation is given by $rI_p$, the limit of $\mathcal{S}$ as the observation time $\tau_m$ approaches infinity is as follows:
\begin{equation}
\mathcal{S}|_{\tau_m \to \infty}=r \mathcal{S}_{\rm cw}=r\alpha_{\rm cw} I_p
\label{eq:lightshift_tauinf}
\end{equation}
From ({\ref{eq:lightshift_tau0}}) and ({\ref{eq:lightshift_tauinf}}), the light shift is shown to be a function that varies from the initial state to reach a steady value with increasing $\tau_m$.
We found that the time-varying function of the light shift is followed by a logistic function of $\tau_m$.
The light shift represented by a general logistic function is as follows:
\begin{equation}
\mathcal{S}=\frac{a_1}{a_2e^{-a_3\tau_m}+1},a_3>0
\label{eq:logistic}
\end{equation}
\noindent
where, $a_1,a_2,a_3$ are values that do not depend on $\tau_m$.
From the initial solution of Eq. ({\ref{eq:lightshift_tau0}}) and the steady-state solution of Eq. ({\ref{eq:lightshift_tauinf}}),
$a_1$ and $a_2$ equal $r\alpha_{\rm cw} I_p$ and $r(\alpha_{\rm cw}C_{\rm lim}^{-1}T I_p+1)-1$, respectively.
When taking the time-dependent behavior of the CPT resonance into account, $a_3$ is equal to the inverse of the pumping time $\tau_p$, which is inversely proportional to the light intensity{\cite{guerandel2007raman}}.
Thus, from Eqs. ({\ref{eq:lightshift_tau0}}), ({\ref{eq:lightshift_tauinf}}), and ({\ref{eq:logistic}}), the light shift behavior in the two-step pulse observation $\mathcal{S}$ can be obtained as follows:
\begin{equation}
\mathcal{S}=\frac{r\alpha_{\rm cw}I_p}{\bigl\{r(\alpha_{\rm cw}C_{\rm lim}^{-1}T I_p+1)-1 \bigr\}e^{-\tau_m/\tau_p}+1}
\label{eq:light shift}
\end{equation}
\noindent
We found that $r$ has an optimal value for suppression of the frequency dependence of the observation time $\tau_m$.
The optimal $r$ is derived as follows:
\begin{equation}
r=\frac{\mathcal{S}|_{\tau_m=0}}{\mathcal{S}_{\rm cw}}=\frac{1}{\alpha_{\rm cw}C_{\rm lim}^{-1}TI_p+1}
\label{eq:condition}
\end{equation}
This means that $r$ has a value such that the light shift of the steady state $\mathcal{S}|_{\tau_m \to \infty}$ is equal to the light shift in the absence of observation.
Under this condition for $r$, the light shift is no longer dependent on $\tau_m$, and the light shift $\mathcal{S}$ equals $\mathcal{S}|_{\tau_m=0}$.

\begin{figure}[]
\centering
\begin{overpic}[scale=0.5]{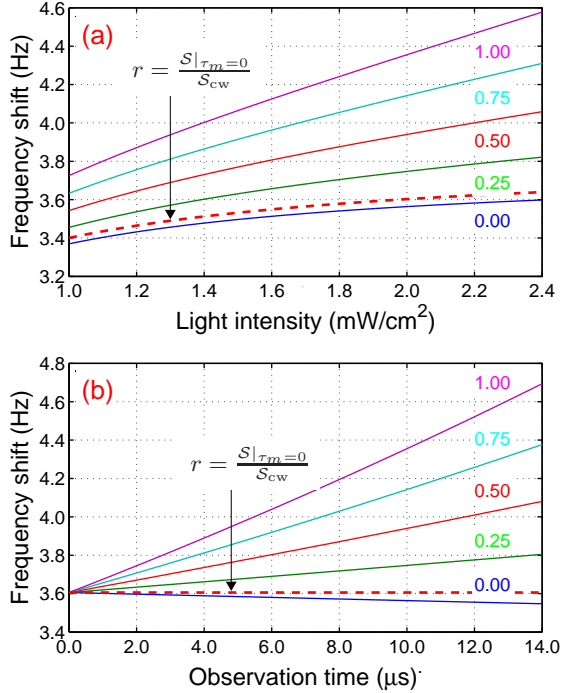}
\put(65,85 ){\colorbox{white}{\small $r=\frac{\mathcal{S}|_{\tau_m=0}}{\mathcal{S}_{\rm cw}}$}}
\put(43,234){\colorbox{white}{\small $r=\frac{\mathcal{S}|_{\tau_m=0}}{\mathcal{S}_{\rm cw}}$}}
\end{overpic}
\caption{(Color online) (a) Light shifts as functions of light intensity under different values of the observation intensity ratio $r$. (b) Light shifts as a function of the observation time under different values of the observation intensity ratio $r$: the parameters are $\alpha_{\rm cw}=30.9$ Hz/(mW/cm$^2$), $C_{\rm lim} = 3.06\times 10^{-3}$ and $\tau_p =99.0$  $\mu$s$\cdot$(mW/cm$^2$)$\times(rI_p)^{-1}$.}
\label{fig:light shift}
\end{figure}

Light shifts as a function of the light intensity under different values of $r$ are shown in Fig. \ref{fig:light shift}(a).
The free evolution time $T$ was 800 $\mu$s, and the observation time $\tau_m$ was set to be 10 $\mu$s.
The results were then calculated numerically from the instantaneous values by using density matrix analysis.
The light shift was determined based on the condition that the dispersion spectrum given by the density matrix analysis is zero.
The zero dispersion was found using Newton's method.
The light shift decreases with decreasing $r$.
Because the pumping rate increases with increasing light intensity, two-step pulse observation with a small $r$ is effective for reduction of the light shift under high light intensity.
The light shift variation is also reduced by the use of a small $r$.
However, the offset shift remains at $r \approx 0.00$.
Because the offset shift is mainly attributed to the free evolution time $T$, a long free evolution time is required to reduce the systematic frequency shift.

Figure \ref{fig:light shift}(b) shows the light shifts as a function of the observation time under different values of $r$.
The light intensity $I_p$ was set to be 2.0 mW/cm$^2$.
The light shift is proportional to $\tau_m$ because the observation time $\tau_m$ is smaller than the pumping time $\tau_p$ in this range.
The frequency variation of the observation time is strongly dependent on $r$ and decreases significantly with decreasing $r$ in the range from 0.25 to 1.00.
Also, the sign of the observation variation changes in the $0.00<r<0.25$ range.
These results indicate that there is an optimal value of $r$ for suppression of the variation in the range of $r$ from 0.00 to 0.25.
The dashed line was plotted by setting $r=\mathcal{S}|_{\tau_m=0}/\mathcal{S}_{\rm cw}$.
$r$ was 0.07 under this condition.
Because the light shift is independent of the observation time setting $r=\mathcal{S}|_{\tau_m=0}/\mathcal{S}_{\rm cw}$, it is not necessary to shorten the observation time to reduce the light shift.
Also, because a small $r$ leads to a low pumping rate, the resonance signal retains its high value, even if we set the observation time to be longer.
Thus, the Ramsey-CPT resonance can be measured without reducing the resonance signal by two-step pulse observation.

Comparison of the results of the numerical calculations performed by density matrix analysis and the estimated values from Eq. (\ref{eq:light shift}) shows that
the relative error of the results is no more than 0.05\%.
The residual of the relative frequency corresponds to the 10$^{-15}$ order, which is very small in the CPT atomic clocks field.
Therefore, the proposed estimation equation is useful for practical purposes.

\begin{figure}[]
\centering
\begin{overpic}[scale=0.5]{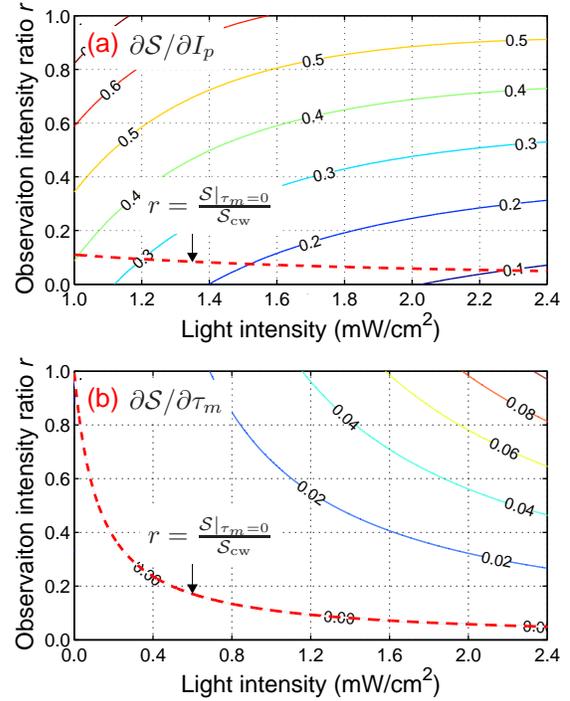}
\put(50,60 ){\colorbox{white}{$r=\frac{\mathcal{S}|_{\tau_m=0}}{\mathcal{S}_{\rm cw}}$}}
\put(50,185){\colorbox{white}{$r=\frac{\mathcal{S}|_{\tau_m=0}}{\mathcal{S}_{\rm cw}}$}}
\put(43,111){\colorbox{white}{$\partial\mathcal{S}/\partial \tau_m$}}
\put(43,245.5){\colorbox{white}{$\partial\mathcal{S}/\partial I_p$}}
\end{overpic}
\caption{(Color online) (a) Contour map of the light shift variation of the light intensity $\partial\mathcal{S}/\partial I_p$ with units of Hz/(mW/cm$^2$). (b) Contour map of the light shift variation of the observation time $\partial\mathcal{S}/\partial \tau_m$ with units of Hz/$\mu$s. The dashed line is a locus that satisfies the condition of Eq. (\ref{eq:condition}).}
\label{fig:map}
\end{figure}
A contour map of the frequency variation of the light intensity is shown in Fig. \ref{fig:map}(a).
Because this variation decreases with increasing light intensity, the smallest variation using the conventional scheme ($r=1.00$) is 0.562 Hz/(mW/cm$^2$) at a light intensity of 2.4 mW/cm$^2$.
The variation with two-step pulse observation decreases with decreasing $r$, and the smallest variation is obtained at $r =0.00$.
The variation at $r=0.00$ was two to seven times smaller than that obtained using the conventional scheme.
The variation reduction effect increases with increasing light intensity.
The smallest variation with the two step pulse observation method is 0.073 Hz/(mW/cm$^2$), which is 7.6 times less than that obtained using the conventional scheme.
The red dashed line indicates a locus of the condition $r=\mathcal{S}|_{\tau_m=0}/\mathcal{S}_{\rm cw}$.
The variation under this locus is higher than that under $r=0.00$; however, the difference is no more than 10\%.
Therefore, a large reduction in the variance can be expected under this locus.

Figure \ref{fig:map}(b) shows the frequency variation of the observation time.
The variation when using the conventional scheme increases significantly with increasing light intensity.
This is because the high light intensity leads not only to a high pumping rate but also to a large frequency difference between $\mathcal{S}_{\rm cw}$ and $\mathcal{S}|_{\tau_m=0}$.
The variation in the observation time is dramatically suppressed under the condition where $r=\mathcal{S}|_{\tau_m=0}/\mathcal{S}_{\rm cw}$.
Because the contour interval is shorter for higher light intensities,
the two-step pulse observation method provides a particular advantage under high light intensity excitation.


\begin{figure}[]
\centering
\includegraphics[scale=0.5]{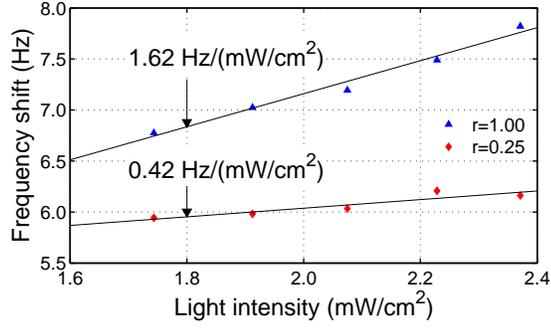}
\caption{(Color online) Experimental results for light shift as a function of the light intensity for various values of $r$.}
\label{fig:experiment}
\end{figure}

Figure \ref{fig:experiment} shows the experimental results for the light shifts as a function of light intensity for various $r$ values.
These results were obtained using a Cs-D$_1$ line vertical-cavity surface-emitting
laser (VCSEL) and a Cs vapor cell.
The light intensity was modulated by an acousto-optical modulator (AOM),
and the two-step pulse shape is formed using the control voltage of the AOM.
Because the absolute resonance signal decreases with decreasing $r$,
the density of the neutral density filter that was located in front of the photodetector was adjusted so that the resonance signal with the two-step pulse observation method was equal to that obtained using the conventional RR scheme.
In this experiment, $r$ was set at 0.25 so that it did not saturate the photodetector voltage during pumping.
The free evolution time was 800 $\mu$s and the observation time was 10 $\mu$s.
The experimental setup is described in detail in Ref. \citen{yano2014theoretical}.
The measured $\alpha_{\rm cw}$ was 28.4 Hz/(mW/cm$^2$).
The solid line is a fitting curve for the experimental data.
The variation obtained using the conventional scheme was 1.62 Hz/(mW/cm$^2$).
The variation obtained with two-step pulse observation ($r=0.25$) is 0.42 Hz/(mW/cm$^2$), which is almost one quarter of that obtained using the conventional scheme.
This reduction is consistent with the calculated results.

In summary, we propose a two-step pulse observation method to enhance the frequency stability of CPT atomic clocks.
The two-step pulse observation method provides a low variation in the light shift without reducing the resonance signal by reducing the steady state repumping.
The calculation results show that there is an optimal value of $r$ for suppression of the frequency variation of the observation time.
We also derive an equation for estimation of the light shift in the RR scheme.
The relative frequency differences between the results obtained using the estimation equation and those from the numerical calculations are of the order of $10^{-15}$, which means that the estimation equation is useful for practical purposes.
The measured light shift is reduced by the two-step pulse observation method, and the reduction in the light shift is reproduced well by the calculations.

\acknowledgement
The authors would like to thank the Ricoh Company, Ltd., for providing us with the Cs-D$_1$ VCSEL.
This work was supported by Grant-in-Aid for JSPS Fellows (no. 26$\cdot$6442).

\end{document}